\documentclass[12pt]{iopart}
\usepackage{graphicx, amssymb}
\begin{document}

\title{2D massless QED  Hall half-integer conductivity and graphene}

\author{A. P\'{e}rez Mart\'{\i}nez}
\address{Instituto de Cibern\'{e}tica Matem\'{a}tica y F\'{\i}sica (ICIMAF) \\
Calle E esq 15 No. 309 Vedado, Havana, 10400, Cuba}
\ead{aurora@icimaf.cu}

\author{E. Rodriguez Querts}
\address{Instituto de Cibern\'{e}tica Matem\'{a}tica y F\'{\i}sica (ICIMAF) \\
Calle E esq 15 No. 309 Vedado, Havana, 10400, Cuba}
\ead{elizabeth@icimaf.cu}

\author{H. P\'{e}rez Rojas}
\address{Instituto de Cibern\'{e}tica Matem\'{a}tica y F\'{\i}sica (ICIMAF) \\
Calle E esq 15 No. 309 Vedado, Havana, 10400, Cuba}
\ead{hugo@icimaf.cu}

\author{R. Gaitan}
 \address{Centro de Investigaciones Te\'oricas, FES-Cuatitl\'an-UNAM}
 \ead{rgaitan@servidor.unam.mx}

\author{S. Rodriguez-Romo}
 \address{Centro de Investigaciones
Te\'oricas, FES-Cuatitl\'an-UNAM}
 \ead{suemi@servidor.unam.mx}

\begin{abstract}
Starting from the photon self-energy tensor in a magnetized medium, the
3D complete antisymmetric form of the conductivity tensor is found in the
static limit of a fermion system $C$ non-invariant under
fermion-antifermion exchange. The massless relativistic 2D fermion
limit in QED is derived by using the compactification along the
dimension parallel to the magnetic field. In the static limit and at
zero temperature the main features of  quantum Hall effect (QHE) are
obtained: the half-integer  QHE and the minimum value proportional
to $e^2/h$ for the Hall conductivity . For typical values of
graphene the plateaus of the Hall conductivity are also reproduced.

\end{abstract}

\maketitle
\section{Introduction}

In 2004, graphene - genuine monolayer of carbon atoms in a honeycomb
array - was obtained experimentally \cite{7, 8}. The theoretical
description of this material was studied with some anticipation
\cite{Wallace}-\cite{Slonczewski} to the experimental results
\cite{McClurePR1956}-\cite{JGonzalezPRB63} and the analogy to the 2D
quantum electrodynamics and  some particular features of this system
were examined in the 1980s \cite{4}-\cite{6}. The experimental
confirmation of its existence obtained as isolated individual
graphene layers leads to increase the theoretical and experimental
studies of its properties with the aim to look for  nanoescale
electronic applications.

Theoretically,  its properties are essentially described by Dirac
massless fermions (electrons) in two dimensions. This system is
¨relativistic¨ in the sense that  the spectra of electrons and holes
can be mimicked as two-dimensional relativistic chiral fermions
where electrons and holes move at velocities $v_F\approx 10^6 m/s$
one hundredth the speed of light \cite{NovoselovNature2005}.

Being the quasiparticles chiral charged massless Dirac fermions, this
has triggered a lot of papers from physicists from high energy
physics (for exhaustive review see Ref. \cite{review2009} and
references therein). From the point of view of high energy physics
graphene could be interesting to test some quantum field theories
and their features in the well known QED, in table top experiments
\cite{Abreu:2010yv} and provides the possibility  to explore exotic
phenomena which could be important for cosmology and astrophysics
\cite{Fradkin}-\cite{Volovik}.

Emerging from theoretical studies of graphene by methods and
technics from Quantum Field Theory (QFT) and condensed matter physics,
one of the main challenges is to gain understanding and matching between both
descriptions and interpretations.

On the other hand the boom in the applications demand the study of
transport properties   \cite{Khodas}, \cite{GusyninPRB2005}, such as
conductivity \cite{GusyninPRB2006}-\cite{JuricicPRB2010} and QHE
\cite{Beneventano:2007fa}-\cite{GusyninPRL2005},  theoretically as
well as experimentally
\cite{JiangaSSC2007}-\cite{NovoselovScience2007}.

The most remarkable result related to the graphene Hall conductivity
is the typical plateau structure for integer quantum filling factors
and the non-zero value of the Hall conductivity when the carrier density
goes to zero \cite{NovoselovNature2005} and these properties have
been  related to the relativistic nature of the graphene dynamics
\cite{Semenoff}-\cite{Luk'yanchuk}.

The effect  has been observed  for densities around $10^{12}
cm^{-2}$,  magnetic fields strength in the range of $15-25 T$ even
at room temperature.

Inspired by these results of QH of graphene we have as a main goal
of the present paper to revisit the relativistic QHE,  especially
the 2D case. Our starting point will be the conductivity tensor
derived in the static limit from the quantum relativistic electron
self-energy tensor in QED in presence of a magnetic field.

As different to  \cite{GonzalezFelipe:1990kq} in which 3D and 2D QHE
was studied with the motivation to apply the results to condensed
matter physics, in the present paper we consider it as model for
graphene-like systems. We start from the simple model of a charged
fermion-antifermion plasma in a magnetic field which is also
interacting  perturbatively with an electromagnetic wave. The study
was done by using  the methods of finite temperature quantum field
theory in the search of its kinetic properties. Two properties are
found for the conductivity tensor. First, in the static limit,
$\omega=0$, $k=0$, it is found that the conductivity tensor is given
by a complete antisymmetric expression for the 3D as well as 2D
cases. Second, the Hall conductivity as a function of the external
magnetic field in the 2D case exhibits the typical plateaus
structure of the integer QHE at zero temperature limit under the
hypothesis of variable number of particles
\cite{Prange}-\cite{Cabo}.
The formalism is a general method, also valid for studying the cases
$\omega \neq 0$, and $k \neq 0$.


The outcomes of the present paper are  interesting due to several
facts:
first, we find the conductivity tensor for both the 3D and 2D cases,
by starting from a massive system, for which we get the massless
limit. (The 2D is obtained as a compactification of the 3D case,
since we are interested in the Hall conductivity of a graphene-like
system). Second, our general expressions can be used for studying
the static as well as the non-static limits. Third, the
non-vanishing of the conductivity in the lowest Landau level
occupancy (leading to the magnetic catalysis) is a consequence of
the $C$-non-invariance of the system. We must also mention that our
basic equations may be used also for studying wave propagation
phenomena, such as Faraday effect; this, however, is out of the
scope of the present paper.


Our results are valid at the one-loop level, for magnetic fields not
very high ($B<<20$ T). Experimental results \cite{Y. Zhang et
al},\cite{Z. Jiang} suggest that higher magnetic fields would demand
to introduce higher loops.

The paper is organized as follows. In section \ref{sec2} we briefly
recall the expression for the charged  fermions  Green function in a
constant magnetic field \cite{Perez Rojas:1979fb},  in section
\ref{sec3}  is obtained the current density and Hall conductivity
starting from the photon self-energy. In Section \ref{sec4} the QH
conductivity is found for the 2D massless fermion system using the
compactification of z-dimension. Finally it is shown the QHE for the
graphene-like system. The  conclusions are drawn  in section
\ref{sec5}.

\section{Green functions for charged fermions in presence of a magnetic field}
\label{sec2}

In this section we outline how to obtain the one~particle Green
function in a medium, e.g., for a system of  charged fermions in
presence of the constant magnetic field at finite temperature
$T=1/\beta$ and density characterized by a chemical potential $\mu$.
It is necessary for the calculation of the photon self-energy tensor
from which the conductivity tensor will be found. In this section we
will take $m \neq 0$ in our expressions and in the next Section we
will take the limit $m = 0$.

The temperature dependent Green function for a gas of charged and
fermions and anti-fermions in a constant external magnetic field
(which we will take parallel to the third axis) $A_\nu =Bx_1
\delta_{2\nu}$, is given by the solution of the Dirac equation
\begin{equation}\label{Dirac_eq}
     [\gamma_\nu (\partial_\nu +ieA_\nu) + m] G(x,x'|A)=\delta(x-x'),
   \end{equation}

\noindent where $\nu=1,2,3,4$, $\partial_4=\partial/\partial_4-\mu$,
and $\mu$ is the chemical potential of the  system. The temperature
Green functions are determined by (\ref{Dirac_eq}) for
$-\beta<x_4<\beta$,  where $\beta=\frac{1}{kT}$,  $k$ is the
Boltzmann constant. For simplicity we use $k=\hbar=c$ and recover
the units at the end of the paper.

According to the established procedure, to obtain the solution of
(\ref{Dirac_eq}) we start from the Fourier transform in time  of the
time dependent Green function, the Fourier parameter $p_0$ is then
continued to the complex value $-ip_4+\mu$, and the resulting
expression, multiplied by $e^{ix_4p_4}$, is summed over the
Matsubara frequencies $p_4=\frac{(2s+1)\pi}{\beta}$, and  $s$ runs
from $-\infty$ to $-\infty$. The time dependent Green function can
be built  from the solutions of the Dirac equation in relativistic
quantum mechanics. We use the following energy eigenfunctions,
according to \cite{Johnson}-\cite{Ahiezer},  the signs $\pm$
correspond to positive and negative energy solutions, respectively

   \begin{eqnarray}\label{eigenfun}
     \phi^{\pm}_{p_2,p_3,n,1}(\mathbf{x}) &=& \large(\frac{\varepsilon_{n,p_3}\pm
     m}{8\pi^2
     \varepsilon_{n,p_3}}\large )^{1/2} e^{ip_2x_2+ip_3x_3}\left(%
\begin{array}{c}
  \psi_{n-1}(\xi) \\
  0 \\
 \frac{ \pm p_{3}}{ (\varepsilon_{n,p_3}\pm
     m)}\psi_{n-1}(\xi)  \\
  \frac{\pm i(2eBn)^{1/2}}{ (\varepsilon_{n,p_3}\pm
     m)}\psi_{n}(\xi) \\
\end{array}%
\right),\\ \nonumber
     \phi^{\pm}_{p_2,p_3,n,-1}(\mathbf{x}) &=&\large(\frac{\varepsilon_{n,p_3}\pm m}{8\pi^2\varepsilon_{n,p_3}}\large)^{1/2} e^{ip_2x_2+ip_3x_3}\left(%
\begin{array}{c}
 0 \\
  \psi_{n}(\xi) \\
   \frac{\mp i (2eBn)^{1/2}}{ (\varepsilon_{n,p_3}\pm
     m)}\psi_{n-1}(\xi) \\
  \frac{\mp  p_{3}}{(\varepsilon_{n,p_3}\pm
     m)}\psi_{n}(\xi)\\
 \end{array}%
\right).
   \end{eqnarray}

\noindent Here the subindex $(p_2,p_3, n, \sigma)$ refers to the
$p_2,p_3$ momenta, to the total quantum number $n$ and to the spin
$\sigma_3$-eigenvalues $\sigma_3=\pm1$, respectively. The energy
eigenvalues are

   \begin{equation}\label{energy}
   \varepsilon_{n,p_3}=\sqrt{p_3^2+ m^2 + 2neB}.
\end{equation}

\noindent These are two fold spin degenerate, except for
$n=0,\sigma_3=-1$, and are also $p_2$ degenerate.

We have written $\xi=\sqrt{eB}(x_1+x_0),$ with  $x_0=p_2/eB$ being
the eigenvalue of the $x_1$ coordinate operator of the center of the
orbit described by the particle) and
\begin{equation}\label{Hermitef}
    \psi_{n}(\xi)=\frac{(eB)^{1/4}}{\pi^{1/4}2^{n/2}(n!)^{1/2}}
    e^{-\frac{\xi^{2}}{2}}H_n(\xi),
\end{equation}

\noindent are the Hermite functions multiplied by $(eB)^{1/4}$.

\noindent The time dependent Green function in the Furry picture is
\begin{equation} \label{greenfunction}
G(\mathbf{x},t,\mathbf{x'},t')=\left \{\begin{array}{cccc}
                                         -i \displaystyle\sum\limits_{q} e^{-i\varepsilon_q(t-t')}{\cal{G}}_{q}^{+}
                                         (\mathbf{x},\mathbf{x'}) &
                                         &\textrm{for}& t>t'\\ \\
                                         i \displaystyle\sum\limits_{q} e^{i\varepsilon_q(t-t')}{\cal{G}}_{q}^{-}
                                         (\mathbf{x},\mathbf{x'}) &  &\textrm{for}& t<t' \\
                                       \end{array} \right.  \hspace{.5cm},
\end{equation}

\noindent where $q$ denotes the set of quantum numbers $(p_2,p_3,
n)$,  $\displaystyle\sum\limits_{q}$ indicates integration on
$p_2,p_3$ and sum over $n=0,1,2,...$ and (the bar means Dirac
adjoint). Here
\begin{eqnarray}\nonumber
{\cal{G}}_{q}^{\pm}(\mathbf{x},\mathbf{x'})&=&\sum_\sigma
\phi^{\pm}_{q,\sigma}(\mathbf{x})\bar{\phi}^{\pm}_{q,\sigma}(\mathbf{x'})\\
\label{Greenf} &=&\frac{e^{ip_2(x_2-x'_2)+ip_3(x_3-x'_3)}}{8\pi^2
\varepsilon_{n,p_3}} \Lambda_q,
\end{eqnarray}
\noindent where
\begin{equation*}
\Lambda_q=\left(%
\begin{array}{cccc}
  C_{n-1,n-1}(\varepsilon_{n,p_3}) & 0 & -D_{n-1,n-1} & -E_{n-1,n} \\
  0 & C_{n,n}(\varepsilon_{n,p_3}) & E_{n,n-1} & D_{n,n} \\
  D_{n-1,n-1} & E_{n-1,n} & C_{n-1,n-1}(-\varepsilon_{n,p_3}) & 0 \\
  -E_{n,n-1} & -D_{n,n} & 0 & C_{n,n}(-\varepsilon_{n,p_3}) \\
\end{array}%
\right),
\end{equation*}
\noindent and
\begin{equation*}
    C_{k,k'}(\varepsilon_{n,p_3})=(\varepsilon_{n,p_3}\pm m)\psi_{k}(\xi)\psi_{k'}(\xi'), \hspace{.5cm}
    D_{k,k'}=\pm p_3\psi_{k}(\xi)\psi_{k'}(\xi'),
\end{equation*}
\begin{equation*}
   E_{k,k'}=\mp i
   (2eBn)^{1/2}\psi_{k}(\xi)\psi_{k'}(\xi').
\end{equation*}

It is understood in (\ref{greenfunction}) that $\psi_{-1}(\xi)\equiv
0$. Taking the Fourier transform in time of (\ref{greenfunction}),
and making the continuation $p_0\rightarrow -ip_4+\mu$, we get for
the $x_4$ Fourier transform of the solution of (\ref{Dirac_eq})
\begin{equation}\label{temGreenf1}
    G(-ip_4+\mu,\mathbf{x},\mathbf{x'})= \frac{{\cal{G}}_{q}^{+}
    (\mathbf{x},\mathbf{x'})}{-ip_4+\mu-\varepsilon_{n,p_3}}+
    \frac{{\cal{G}}_{q}^{-}
    (\mathbf{x},\mathbf{x'})}{-ip_4+\mu+\varepsilon_{n,p_3}}.
\end{equation}

\noindent After multiplication by $e^{ip_4 x_4}$ and summation over
$p_4$ we have the following expression for the temperature dependent
Green function:
\begin{eqnarray} \label{tempGreenf1}\nonumber
 G(\mathbf{x},\mathbf{x'})&=& \displaystyle \sum_q\left[(n_e(\varepsilon_{n,p_3})-1)e^{-(\varepsilon_{n,p_3}-\mu)(x_4-x'_4)}{\cal{G}}_{q}^{+}
    (\mathbf{x},\mathbf{x'})\right.  \\ \nonumber & & \hspace{2.1cm} \left.-n_p(\varepsilon_{n,p_3})e^{(\varepsilon_{n,p_3}+\mu)(x_4-x'_4)}{\cal{G}}_{q}^{-}
    (\mathbf{x},\mathbf{x'})\right]\hspace{.2cm}
                                         \textrm{for } x_4>x'_4,\\
                                         & & \\\nonumber
    G(\mathbf{x},\mathbf{x'})&=& \displaystyle \sum_q
  \left[n_e(\varepsilon_{n,p_3})e^{-(\varepsilon_{n,p_3}-\mu)(x_4-x'_4)}{\cal{G}}_{q}^{+}
    (\mathbf{x},\mathbf{x'})\right. \\ \nonumber
    & & \hspace{1cm} \left.-(n_p(\varepsilon_{n,p_3})-1)e^{(\varepsilon_{n,p_3}+\mu)(x_4-x'_4)}{\cal{G}}_{q}^{-}
    (\mathbf{x},\mathbf{x'})\right] \hspace{.2cm}  \textrm{for } x_4<x'_4,
    \end{eqnarray}

\noindent where
\begin{eqnarray}
 \nonumber
  n_e(\varepsilon_{n,p_3}) &=& \frac{1}{1+e^{(\varepsilon_{n,p_3}-\mu)\beta}}, \\
  n_p(\varepsilon_{n,p_3}) &=& \frac{1}{1+e^{(\varepsilon_{n,p_3}+\mu)\beta}},
\end{eqnarray}

\noindent are the mean number of fermions and anti-fermions,
respectively.

\noindent Later we will be interested in the massless and 2D limits
of the Green function (\ref{tempGreenf1}).

\section{Current density and conductivity from  the photon self-energy}
\label{sec3}

This section is devoted to obtain the general expression of the
conductivity tensor for the quantum relativistic  fermion plasma
which is linear in the perturbative electromagnetic field $A_{\mu}$.
In Euclidean variables we have the Maxwell equations $D_{0\mu
\nu}^{-1}A_\nu=j_{\mu}(A)$, where $D_{0\mu
\nu}^{-1}=(\partial_{\lambda}^{2}\delta_{\mu \nu}- \partial_\mu
\partial_\nu)$

Our starting point will be the linear term in $A_\mu$ in the
expansion of $j_{\mu}(A)$ in powers of $A_\lambda$, the coefficient
being the photon self-energy tensor. The current density  can be
written  as

\begin{equation}
j_i=\pi_{i \nu}A_{\nu}=Y_{ij}E_j,\,\,\, \nu=1,2,3,4, \,\,\,
i=1,2,3,\label{corriente}
\end{equation}

\noindent where $E_j=i(\omega A_j-k_jA_0)$ is the electric field of
the electromagnetic wave, $A_4=iA_0$, $k_4=i\omega$ and
$Y_{i,j}=\pi_{ij}/i\omega$ is the complex conductivity tensor or
admittivity, and the third term in (\ref{corriente}) comes from the
second one by using the four-dimensional transversality of $\pi_{\mu
\nu}$, $\pi_{\mu \nu}k_\nu =0$, due to gauge invariance.

We are interested in the real conductivity which can be expressed in
terms of the imaginary part of the photon self-energy and the
frequency as $\sigma_{ij}=\mathrm{Im} \pi_{ij}/\omega$ \cite{Perez
Rojas:1979fb}.

The components of the photon self-energy tensor in the case of non
zero temperature and density and in presence of the external
magnetic field $B$ can be obtained from

\begin{equation}\label{polop}
    \pi_{\nu\mu}(x,y)=e^2 Tr\int \gamma_\mu G(x,z)\Gamma_\nu
    (z,y',y)G(y',x)d^{3}z d^{3}y'.
\end{equation}

\noindent $ G(y',x)$ is the Green function of the charged fermions
obtained in section I and $\Gamma_\nu$ is the vertex function which
in one loop approximation has the form $\Gamma_\nu=
\gamma_\nu\delta(z-y')\delta(z-y)$.

In  \cite{Perez Rojas:1979fb}-\cite{Perez Rojas:1982} it was
obtained the expression (\ref{polop}) for electron positron plasma.
Details of the calculations can be seen there. In that paper  it was
shown that $\pi_{\mu\nu}$ can be expressed as a linear combination
of the six four-dimensional transverse tensors, four of them
symmetric and two antisymmetric with respect to the indices
$\mu,\nu$, $\pi_{\mu\nu}=\sum_{i=1}^6\pi^{i}\Phi^{i}_{\mu\nu}$. The
scalar coefficients $\pi^{(i)}$ in that expansion are expressed in
terms of six scalars functions $p,t,s,q,r$ and $v$ \cite{Perez
Rojas:1979fb}(after the dimensional reduction $3 + 1 \to 2 + 1$ the
scalars $q$ and $v$ vanish and the number of independent scalars is
reduced to four, $p, t, s, r$). By studying the analytic properties
of $\pi_{\mu\nu}$ it was proved that the imaginary components of its
symmetric part are due to the singularities produced by the photon
absorptive processes (excitations of the electrons and positrons and
pair creation), whereas the imaginary part of its antisymmetric
terms  is connected with the interaction of the net charge  and
current of the electron-positron system with the external magnetic
field. The first mechanism contributes to Ohm conductivity whereas
the second, to Hall conductivity.

Let us note that all these arguments can be extended to our present
calculation and the contribution to the current density $j_i$ in the
equation (\ref{corriente}) due to conductivity can be written then
in the general form  as

\begin{equation}
j_i=\sigma_{ij}^0E_j + (E\times \mathcal{S})_i,  \label{corriente1}
\end{equation}

\noindent where $\sigma_{ij}^0= \mathrm{Im}\pi_{ij}^S/\omega$ and
$\mathcal{S}_{i}=\frac{1}{2}\epsilon^{ijk}\sigma_{jk}^H$ is a
pseudovector associated to  $\sigma_{jk}^H=  \mathrm{Im}
\pi_{ij}^A/\omega$, $\pi_{ij}^S$, $\pi_{ij}^A$ are the symmetric and
antisymmetric parts of the spatial photon self-energy. The first
term in (\ref{corriente1}) corresponds to the Ohm current and the
second is the Hall current E is the electric field  corresponding to
the electromagnetic wave eigenmodes of  $\pi_{\mu\nu}$. Let us
consider the specific case when the electric field $\mathbf{E}$ is
due to a transverse mode propagating along the magnetic field
$\mathbf{B}$, and thus $\mathbf{E}$ is perpendicular to $\mathbf{B}$
and does not depend on the components $k_1, k_2$
$(\mathbf{B}\parallel k_3)$. The expression for the conductivity,
according to \cite{GonzalezFelipe:1990kq}, is

\begin{equation}
\sigma_{ij}=  \sigma^0_{3D}\delta_{ij} + \epsilon^{ij}\sigma^H_{3D},
\label{conductividad}
\end{equation}

\noindent where $\epsilon^{ij}$ is the antisymmetric $2\times 2$
unit tensor, $\epsilon^{12}=-\epsilon^{21}=1$ and $\sigma^0_{3D}=
\mathrm{Im} t/\omega$, $\sigma^H_{3D}= \mathrm{Im} r/\omega$ and the
expressions for the scalar quantities $\mathrm{Im} t$, $\mathrm{Im}
r$  were found in \cite{GonzalezFelipe:1990kq}.
 Let us take the limit $m \to 0$, that is, let us consider  a version of QED in the fermion massless limit. This would require a more detailed discussion, since chiral invariance appears
 and also  the Feynman diagrams and renormalizability must be discussed in the new context. We assume that limit as satisfactory and proceed with its consequences. We have now
\begin{equation}\label{t}
\mathrm{Im} t=\frac{e^3B}{4\pi}\sum\limits_{n,n'=0}^{\infty}
\frac{\Upsilon(\delta_{n,n'-1}+\delta_{n-1,n'})[N^+(\epsilon_q)-N^+(\epsilon_q+\omega)]}{\sqrt{(z_1+2eB
(n-n'))^2+ 4z_1\varepsilon_{n,0}}},
\end{equation}

\noindent where
$$\Upsilon=z_1+2eB(n+n'),\hspace{.5cm}z_1= (k_3^2-\omega^2),\hspace{.5cm}
N^+(\epsilon_q)=n_e(\epsilon_q)+n_p(\epsilon_q), \hspace{.5cm}$$
$$\epsilon_q=\frac{-\omega z_1+|k_3|\sqrt{(z_1+2eB(n-n'))^2+ 4z_1\varepsilon_{n,0}}}{2z_1}, \,\,\,
\varepsilon_{n,0}=\sqrt{2eBn}$$ and
\begin{equation}\label{r}
\mathrm{Im}
r=\frac{e^3B\omega}{2\pi}\sum\limits_{n,n'=0}^{\infty}(\delta_{n,n'-1}-\delta_{n-1,n'})\int
dp_3\frac{\Upsilon}{|Q|^2}(n_e(\varepsilon_{n,p_3})-n_p(\varepsilon_{n,p_3})),
\end{equation}

$$|Q|^2=[2 p_3k_3-z_1+2eB(n-n')]^2-4\omega^2(2eBn). $$

\noindent It is easy to prove that in the static limit $\omega=0$,
$k_3=0$, the conductivity tensor becomes completely antisymmetric
($\sigma^0_{3D}=0$), because the term $ \mathrm{Im} t/\omega$
vanishes in that limit \cite{GonzalezFelipe:1990kq} and the
conductivity is $\sigma_{ij}=\sigma^{H}_{3D}\epsilon^{ij}$. However,
the Hall conductivity is non zero in the same limit and it has the
form

\begin{equation}\label{3DHall}
  \sigma^H_{3D}=\frac{e^2}{4\pi^2}\sum_{n=0}^{\infty}\alpha_n\int_{-\infty}^{\infty}dp_3(n_e(\varepsilon_{n,p_3})-n_p(\varepsilon_{n,p_3})),\hspace{1cm}
  \alpha_n=2-\delta_{0n}.
\end{equation}

\noindent The expression (\ref{3DHall}) is valid for the massive as
well as for the massless limit. If the system is $C$-invariant, as
in a neutral gas of electrons and positrons, $\mu=0$ and it implies
$\sigma^H_{3D}=0$. No Hall current is excited. But if the system is
not $C$-invariant, as it happens in ordinary matter composed from
charged baryons and leptons, this is not the case (see below). At
zero temperature and for $\mu>0$ ($\mu<0$) the contribution of
antifermions (fermions) is zero, the fermion  gas is completely
degenerate, and $\sigma^H_{3D}$ takes in the massless case the form

\begin{eqnarray}\nonumber
  \sigma^H&=&\pm\frac{e^2}{4\pi^2}\sum_{n=0}^{n_{\mu}}\alpha_n\int\limits_{-\infty}^{\infty}dp_3
  \theta(\mu\mp\varepsilon_{n,p_3})\\\label{Hall}
  &=&\pm\frac{e^2}{4\pi^2}\sum_{n=0}^{n_{\mu}}\alpha_n
 \sqrt{\mu^2-2eBn}.
\end{eqnarray}

\noindent where $n_{\mu}=I( \mu^2/2eB )$, and $\theta(x)$,  $I(x)$
are the step and the integer functions, respectively. Obviously, for
$\mu <\sqrt{2eB}$ only the lowest Landau level (LLL) is occupied and
$\sigma^H=\pm\frac{e^2}{4\pi^2}\mu$ does not depend explicitly on
$B$. However, if the density $N$ is fixed, $\mu$ depends both on $N$
and on $B$ . But if  $\mu$ is fixed, the condition of occupying only
the LLL is always achieved by choosing $B$ large enough. Actually,
this is valid even in the massive case, as is seen if in the
previous inequality $\mu$ is replaced by $\mu'=\sqrt{\mu^2-m^2c^4}$
as the effective chemical potential.

We must  emphasize that the properties of
(\ref{3DHall})-(\ref{Hall}) come from the scalar $r$ which as well
as other components of $\pi_{\mu\nu}$ emerge from the relativistic, charge conjugation, parity and time reversal CPT and gauge invariance of  QED.

Our next step will be to obtain the relativistic 2D Hall
conductivity which describe  a graphene-like system.

\section{2D massless fermion conductivity}
\label{sec4}

Let us describe  the 2D massless  QED behavior as an effective 2D
theory, which is obtained from the 3D one, after dimensional
compactification. To that end we assume that the coordinate $z$ is
compact with a certain compactification length $L_3$; that is, $z$
varies in the interval $0\leq z \leq L_3$, and points $z = 0$ and $z
= L_3$ are identified. The momentum along the $z$ direction will be
then quantized $p_3=2\pi s/L_3$, $s=0,1,2,...$, and below the energy
scale $E_0=2\pi /L_3$ only zero modes with $s= 0$ (corresponding to
$p_3=0$) are relevant, i.e. the theory  is effectively
two-dimensional. In this reduced space, the external magnetic field
$B$ behaves like a pseudo-scalar. In this section we will return to
units $c$, $h=2\pi \hbar$, starting from equation (\ref{2DHall0}).

The mentioned zero modes with $p_3=0$ satisfy then the reduced Dirac
 equation
\begin{equation}     \gamma_\nu (\partial_\nu +ieA_\nu)
G(x,x'|A)=\delta(x-x'),
   \end{equation}
  where $\nu=1,2,4$, and the  corresponding energy eigenfunctions
  are the $m=0$, $p_3 =0$ limits of (\ref{eigenfun})
   \begin{eqnarray}\label{2Deigenfun}
     \phi^{\pm}_{p_2,0,n,1}(\mathbf{x}) &=& \frac{1}{2\pi^{1/2}L_3} e^{ip_2x_2}\left(%
\begin{array}{c}
  \psi_{n-1}(\xi) \\
  0 \\
0  \\
  \pm i(2eBn)^{1/2}\varepsilon_{n,0}^{-1}\psi_{n}(\xi) \\
\end{array}%
\right),\\ \nonumber
     \phi^{\pm}_{p_2,0,n,-1}(\mathbf{x}) &=& \frac{1}{2\pi^{1/2}L_3}  e^{ip_2x_2}\left(%
\begin{array}{c}
 0 \\
  \psi_{n}(\xi) \\
   \mp i(2eBn)^{1/2}\varepsilon_{n,0}^{-1}\psi_{n-1}(\xi) \\
 0\\
 \end{array}%
\right).\label{Diraceq2D}
\end{eqnarray}

  The eigenvalues $\sigma_3=\pm1$ are now interpreted, from the 2D point of view, as the pseudo spin quantum numbers, 
     and the energy eigenvalues are simply
   \begin{equation}\label{2Denergy}
   \varepsilon_{n,0}=\sqrt{2neB}.
\end{equation}

Let us remark that as different from the $3+1$ QED in which
there is only one $4 \times 4$ irreducible representation for the
Dirac $\gamma$-matrices, in $2 + 1$ dimensions \cite{Gusynin2008}
there are two $2 \times 2$ inequivalent irreducible representations
of $\gamma$-matrices (which in graphene correspond to two points of
the Fermi surface, $K$ and $K^{\prime}$ describing respectively
states on sublattices ($A$ , $B$) of the hexagonal lattice).
Therefore, the dimensional reduction we have made correspond to the
sum of these two $2 \times 2$ irreducible representations. However,
from (\ref{Diraceq2D}), which is the $2D + 1$ limit of
(\ref{eigenfun}), we see that to each spin state $\pm 1$ correspond
two (obviously inequivalent) positive and negative energy
eigenstates $\phi^{\pm}$.

\noindent All the expressions obtained previously from the photon
self-energy are translated immediately to the 2D case. For instance,
in the static limit $\omega=0$ ($k_3=0$ by assumption), and to
obtain the zero mode  3D Hall conductivity  we should substitute in
(\ref{r}) $p_3=2\pi s/L_3$,
 replace the integral over $p_3$ by a sum over the integers $s$
and retain only the  $s=0$ term. Then, the corresponding zero mode
2D Hall conductivity $\sigma^H_{2D}\equiv L_3\sigma^H_{3D}$ is
\begin{eqnarray}\nonumber
\sigma^H_{2D}&=&\frac{L_3\mathrm{Im} r}{\omega}\\
&=&
\frac{e^2}{h}\sum_{n=0}^{\infty}\alpha_n(n_e(\varepsilon_{n,0})-n_p(\varepsilon_{n,0})).\label{2DHall0}
\end{eqnarray}

At zero temperature, for positive (negative) chemical potential
$\mu>0$ ($\mu<0$), the contribution of antifermions (fermions) is
zero, the fermion (antifermion) gas is completely degenerate. Taking
into account the spectrum energy (\ref{2Denergy}) and by considering
that the chemical potential is $n_\mu<|\mu^2|/2eB<n_\mu+1$, we
obtain for $\sigma^H_{2D}$

  \begin{equation}\label{2DHallCond}
\sigma^H_{2D}=\left\{\begin{array}{cccc}
                                   \displaystyle  \frac{e^2}{h}\sum\limits_{n=0}^{n_\mu}\alpha_n\theta
                                   (\mu-\varepsilon_{n,0})
                                   =  2\frac{e^2}{h}\left(n_\mu+\frac{1}{2}\right)&
                                   &\textrm{for}& \mu>0\\ \\
                                   -\displaystyle \frac{e^2}{h} \sum\limits_{n=0}^{n_\mu}\alpha_n\theta
                                   (\mu+\varepsilon_{n,0})
                                   =- 2\frac{e^2}{h}\left(n_\mu+\frac{1}{2}\right)&
                                   &\textrm{for}& \mu<0 \\
                                   \end{array} \right.
                                   \hspace{.5cm}. \label{sigmah}
\end{equation}

 The expression (\ref{2DHallCond})
describes the anomalous Hall conductivity of the 2D relativistic
massless fermions and antifermions.

\subsection{Thermodynamical potential for 2D fermions}

The Hall conductivity obtained before can be derived from the 2D
thermodynamical potential which  is obtained from the 3D expression
\cite{hugo3} and has the following form

\begin{equation}\label{Omega}
\Omega_{2D}= -\frac{eB}{h c\beta}\sum_{n=0}^{\infty} \alpha_n \ln
(1+e^{-(\varepsilon_{n,0}-\mu)\beta})(1+e^{-(\varepsilon_{n,0}+\mu)\beta}).
\end{equation}

The particle density  is $N=-\frac{\partial \Omega}{\partial \mu}$,
for the 2D system at zero temperature  has the form

\begin{equation}\label{density2D}
N_{2D}=\frac{eB}{h c}\sum_{n=0}^{\infty}\alpha_n\theta
(\mu-\varepsilon_{n,0}).
\end{equation}

\noindent The expression for the density (\ref{density2D}) shows a
non-zero value at the LLL. This leads to the so-called magnetic
catalysis (MC) phenomenon, a chiral symmetry breaking determined  by
the dimensional reduction of the charged system in a magnetized
medium \cite{GorbarPRB2008}.  Consequently the absolute value of the
2D Hall conductivity (\ref{2DHallCond}) has a minimum value of
order $2\frac{e^2}{h}$ for small $\mu >T \to 0$ such that $0<\frac{\mu^2}{2eB}<1$.

In presence of an uniform electric field  it has been proved in \cite{Giaconni} that the minimum quantum conductivity in graphene is $4e^2/h$.

It is possible to rewrite the Hall conductivity (\ref{2DHallCond})
in terms of the particle density as given by the well-known
classical expression  \cite{Kittel}  as

\begin{equation}\label{2DHallCond2}
  \sigma^H_{2D}=\frac{ec  N_{2D}}{B},
\end{equation}

\noindent which is a  result obtained for 3D Hall conductivity but
we take now $N$ as the 2D density, (number of particles per unit
area).

This result agree with the calculations presented in
\cite{Beneventano:2007fa} where the authors have started from the
effective action of the 2D fermion gas in  presence of a magnetic
field which is equivalent to the treatment  of the thermodynamical
potential  as a way to confirm that our results, obtained through
the photon self energy, are in agreement with the more usual form of
dealing with the Hall conductivity.

Previously we have dealt with a C-invariant system of
particles-antiparticles (as electrons and positrons) being their
chemical potentials $\mu_e =\mu$, $\mu_p=-\mu$ in the frame of QED
at finite temperature. As we want to understand our results as a
graphene-like system, we note that for graphene the electron-hole
subsystem behaves also as C-invariant \cite{Katsnelson}. Thus, we must replace the positron distribution $n_p$ by the hole distribution $n_h$ in (\ref{2DHall0})). We get then  at $T\neq 0$, that if $\mu=0$,
the carrier density $N_{e,h}=\frac{e B}{h
c}\sum_{n=0}^{\infty}\alpha_n(n_e-n_h)=0$ and in consequence
$\sigma^{H}_{2D} = 0$,  the Hall conductivity goes smoothly to
zero when the chemical potential swaps from negative to positive
values \cite{GusyninPRB2006}.

\subsection{Graphene half-integer Hall conductivity}

 Expressions (\ref{2DHallCond}), (\ref{2DHallCond2})  can
also describe a graphene like half-integer Hall conductivity  by
substituting for the values of the typical parameters of graphene.
In addition to the the spin degeneracy, an extra factor two comes
from the sublattice-valley degeneracy in graphene.
Also the substitution $c\rightarrow v_F^2/c$, where $v_F$ is the
Fermi velocity must be made in the energy eigenvalue equation and in
other corresponding places). Finally we get the well known
expression for the graphene Hall conductivity (measured in c.g.s.
units) as

\begin{equation}
  \sigma^H=\pm 4 \frac{e^2}{h} (n_g+1/2), \hspace{1cm} n_g=I(\frac{\mu^2 c}{2eB\hbar
  v_F^2}).\label{grcond}
\end{equation}

 In Figs. 1 and 2 we have plotted the Hall conductivity as a
function of magnetic field and density for typical values of the
experiments where QHE has been observed. As can be seen, the
degenerate limit ($T=0$) describes quite well the behavior for $T\ll
T_{F}$ which, is fulfilled for graphene-like system where the effect
is observed at room temperatures (300 K) with densities  of
$10^{12}cm^{-2}$.

\begin{figure}[h!]
\begin{center}
\includegraphics[width=0.8\textwidth]{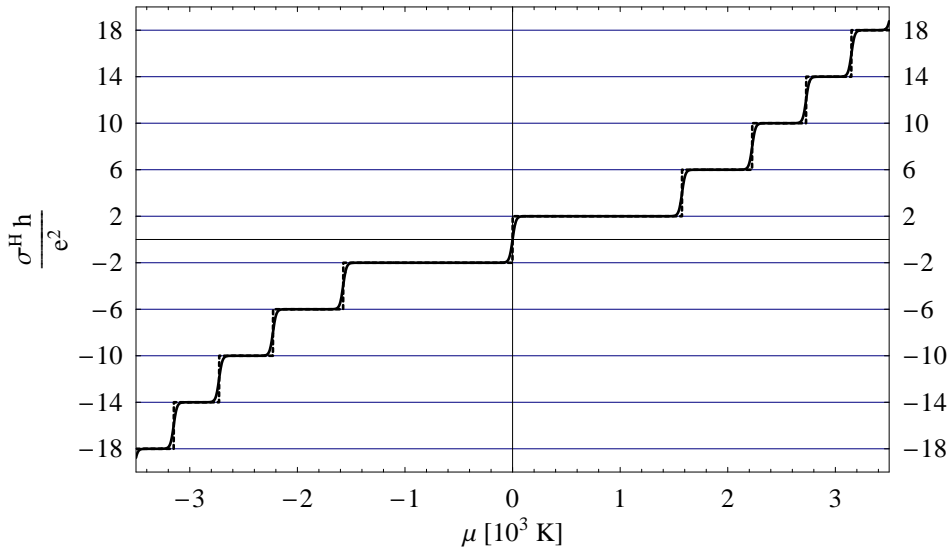}
\end{center}
\caption{Graphene-like Hall conductivity $\sigma^H$ (measured in $e^2/h$
 units) as a function of chemical potential $\mu$ for  $B=1.4*10^5$ G,  $T=0$ K (dashed line) and $T=15$ K (continuous line). We use $v_F=1.0\times10^{8}$ $cm/s$, which leads to an energy scale  (given in units of temperature) $\frac{eB\hbar
 v_F^2}{c} \rightarrow$ $\frac{eB\hbar v_F^2}{ck_B^2}$(K$^2$) $=8.858\frac{K^2}{G}$B(G).}
 \end{figure}
\begin{figure}[h]
\begin{center}
\includegraphics[width=0.8\textwidth]{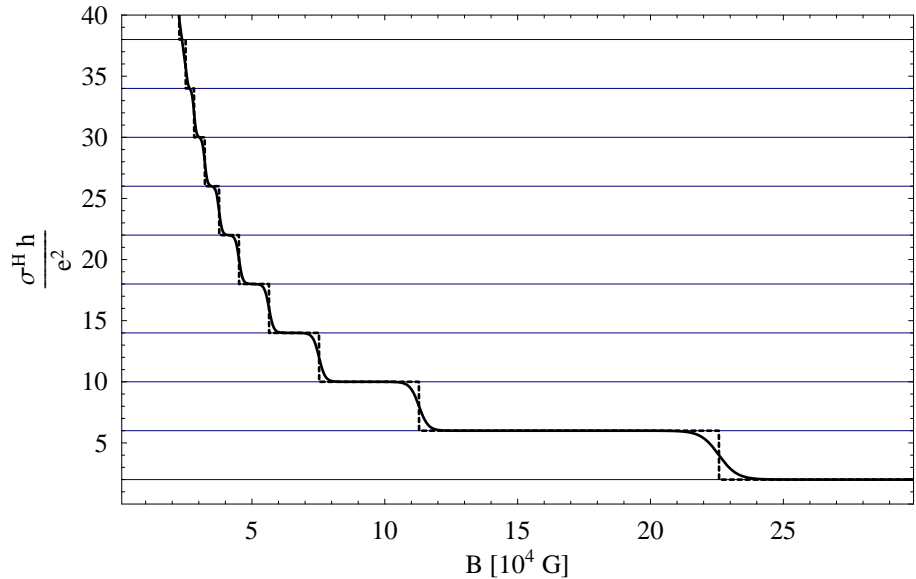}
\end{center}
 \caption{Graphene-like Hall conductivity $\sigma^H$ (measured in $e^2/h$
 units) as a function of magnetic field  $B$ for  $\mu=2\times10^{3}$ K, $T=0$ K (dashed line) and $T=15$ K (continuous line).}
\end{figure}

\section{Conclusions}
\label{sec5} To summarize, we have presented  some kinetic
properties of the 3D quantum relativistic massless fermion gas in
presence of the magnetic field, due to the $C$ non-invariance of the
system, and we have obtained the conductivity tensor in the static
limit showing that it is completely antisymmetric, since the Ohm
conductivity vanish $\sigma_{0}=0$ and the Hall conductivity remains
as different from zero at the limit of zero temperature. With the
aim to study  2D relativistic system we have introduced an ansatz:
the dimensional compactification  of $z$-dimension  allowing to
obtain the  Hall conductivity of the system  showing  the plateaus.
From this framework it is possible to obtain the graphene-like
half-integer Hall conductivity, if an extra degeneracy factor two,
due to the sublattice-valley of graphene, is introduced. Our results
are in agreement with the results of
\cite{NovoselovNature2005}-\cite{Beneventano:2007fa}  where the Hall
conductivity was obtained. The absolute value of graphene
conductivity, as well as the 2D relativistic Hall conductivity,
exhibits a minimum value of order $e^2/h$. This phenomenon is
connected to the magnetic catalysis mechanism \cite{GorbarPRB2008}.

\ack{
We are grateful to A Cabo Montes de Oca and M Perez Maldonado for very useful comments and suggestions. The authors APM and ERQ acknowledge CIT-FES UNAM Cuatitlan Izcalli for the hospitality. APM, HPR and ERQ have been supported by Ministerio de Ciencia, Tecnologia y Medio Ambiente under the grant CB0407, and they also thank the ICTP Office of External Activities for its support through NET-35. S.R.R and R.G thanks to the support of the project  PAPIIT No IN117611-3.  \\}


\end{document}